\documentclass{jpsj-suppl}
\usepackage{txfonts} %Please comment out this line unless the txfonts package is availabe in your LaTeX system.

\usepackage[usenames]{color}

\newcommand{\rE}{\ensuremath{r_{\mathrm E}}}

\newcommand{\rp}{\ensuremath{\rE({\mathrm p})}}
\newcommand{\rd}{\ensuremath{\rE({\mathrm d})}}

\newcommand{\mup}{\ensuremath{\mu \mathrm{p} }}
\newcommand{\mud}{\ensuremath{\mu \mathrm{d} }}
\newcommand{\muHet}{\ensuremath{(\mu\,^3 \mathrm{He})^+ }}
\newcommand{\muHef}{\ensuremath{(\mu\,^4 \mathrm{He})^+ }}
\newcommand{\mum}{\ensuremath{\mu \mathrm{m} }}
\newcommand{\mus}{\ensuremath{\mu \mathrm{s} }}
\newcommand{\Htwo}{\ensuremath{\mathrm H_{2}}}
\newcommand{\Dtwo}{\ensuremath{\mathrm D_{2}}}

\newcommand{\mumi}{\ensuremath{\mu^- }}

\newcommand{\DLineOne}{\ensuremath{2S_{1/2}^{F=3/2} \rightarrow 2P_{3/2}^{F=5/2}}}
\newcommand{\DLineTwo}{\ensuremath{2S_{1/2}^{F=1/2} \rightarrow 2P_{3/2}^{F=3/2}}}
\newcommand{\DLineThree}{\ensuremath{2S_{1/2}^{F=1/2} \rightarrow 2P_{3/2}^{F=1/2}}}

\def\iMPQ{1}
\def\iMainz{2}
\def\iLKB{3}
\def\iCOI{4}
\def\iIFSW{5}
\def\iLISBON{6}
\def\iAVEIRO{7}
\def\iYALE{8}
\def\iPSI{9}
\def\iDG{10}
\def\iETHZ{11}
\def\iFR{12}
\def\iTAIWAN{13}
\def\iPRINCE{14}

\title{Laser Spectroscopy of Muonic Atoms and Ions}

\author{
Randolf~\textsc{Pohl},$^{\iMPQ,\iMainz}$
Fran\c{c}ois~\textsc{Nez},$^{\iLKB}$
Luis~M.~P.~\textsc{Fernandes},$^{\iCOI}$
Marwan~\textsc{Abdou Ahmed},$^{\iIFSW}$
Fernando~D.~\textsc{Amaro},$^{\iCOI}$
Pedro~\textsc{Amaro},$^{\iLISBON}$
Fran\c{c}ois~\textsc{Biraben},$^{\iLKB}$
Jo\~{a}o~M.~R.~\textsc{Cardoso},$^{\iCOI}$
Daniel~S.~\textsc{Covita},$^{\iAVEIRO}$
Andreas~\textsc{Dax},$^{\iYALE,\iPSI}$
Satish~\textsc{Dhawan},$^{\iYALE}$
Marc~\textsc{Diepold},$^{\iMPQ}$
Beatrice~\textsc{Franke},$^{\iMPQ}$
Sandrine~\textsc{Galtier},$^{\iLKB}$
Adolf~\textsc{Giesen},$^{\iIFSW,\iDG, \dagger}$
Andrea~L.~\textsc{Gouvea},$^{\iCOI}$
Johannes~\textsc{G{\"o}tzfried},$^{\iMPQ}$
Thomas~\textsc{Graf},$^{\iIFSW}$
Theodor~W.~\textsc{H{\"a}nsch},$^{\iMPQ, \spadesuit}$
Malte~\textsc{Hildebrandt},$^{\iPSI}$
Paul~\textsc{Indelicato},$^{\iLKB}$
Lucile~\textsc{Julien},$^{\iLKB}$
Klaus~\textsc{Kirch},$^{\iETHZ,\iPSI}$
Andreas~\textsc{Knecht},$^{\iPSI}$
Paul~\textsc{Knowles},$^{\iFR, \S}$
Franz~\textsc{Kottmann},$^{\iETHZ}$
Julian~J.~\textsc{Krauth},$^{\iMPQ}$
Eric-Olivier~\textsc{Le~Bigot},$^{\iLKB}$
Yi-Wei~\textsc{Liu},$^{\iTAIWAN}$
Jos\'{e}~A.~M.~\textsc{Lopes},$^{\iCOI, \#}$
Livia~\textsc{Ludhova},$^{\iFR, \diamondsuit}$
Jorge~\textsc{Machado},$^{\iLISBON}$
Cristina~M.~B.~\textsc{Monteiro},$^{\iCOI}$
Fran\c{c}oise~\textsc{Mulhauser},$^{\iFR,\iMPQ, \odot}$
Tobias~\textsc{Nebel},$^{\iMPQ}$
Paul~\textsc{Rabinowitz},$^{\iPRINCE}$
Joaquim~M.~F.~\textsc{dos~Santos},$^{\iCOI}$
Jos{\'e}~Paulo~\textsc{Santos},$^{\iLISBON}$
Lukas~A.~\textsc{Schaller},$^{\iFR}$ 
Karsten~\textsc{Schuhmann},$^{\iETHZ,\iDG,\iPSI}$
Catherine~\textsc{Schwob},$^{\iLKB}$
Csilla~I.~\textsc{Szabo},$^{\iLKB, \ddagger}$
David~\textsc{Taqqu},$^{\iPSI}$
Jo\~{a}o~F.~C.~A.~\textsc{Veloso},$^{\iAVEIRO}$ and
Andreas~\textsc{Voss},$^{\iIFSW}$
Birgit~\textsc{Weichelt},$^{\iIFSW}$
Aldo~\textsc{Antognini}$^{\iMPQ,\iETHZ,\iPSI}$\\[1ex]
The CREMA Collaboration\\
}

\inst{
{$^{\iMPQ}$Max--Planck--Institut f{\"u}r Quantenoptik, 85748 Garching,
  Germany.}\\ 
{$^{\iMainz}$Johannes Gutenberg-Universit{\"a}t Mainz,
QUANTUM, Institut f{\"u}r Physik \& Exzellenzcluster PRISMA,
55099 Mainz, Germany.} \\
{$^{\iLKB}$Laboratoire Kastler Brossel, UPMC-Sorbonne Universit\'es, CNRS, }\\ 
{ENS-PSL Research University, Coll\`{e}ge de France,
75005 Paris, France.}\\
{$^{\iCOI}$LIBPhys, Department of Physics, University of Coimbra,
  3004-516 Coimbra, Portugal.}\\
{$^{\iIFSW}$Institut f{\"u}r Strahlwerkzeuge, Universit{\"a}t Stuttgart,
-  70569 Stuttgart, Germany.}\\
{$^{\iLISBON}$ Laborat\'orio de Instrumenta\c{c}\~ao, Engenharia Biom\'edica e F\'isica da Radia\c{c}\~ao
(LIBPhys-UNL),~Departamento de F\'isica, Faculdade~de~Ci\^{e}ncias~e~Tecnologia,~FCT,~Universidade Nova de Lisboa,~2829-516 Caparica, Portugal.}\\
{$^{\iAVEIRO}$I3N, Departamento de F\'isica, Universidade de Aveiro,
  3810--193 Aveiro, Portugal.}\\
{$^{\iYALE}$Physics Department, Yale University, New Haven,
  CT 06520--8121, USA.}\\
{$^{\iPSI}$Paul Scherrer Institute, 5232 Villigen--PSI, Switzerland.}\\
{$^{\iDG}$Dausinger \& Giesen GmbH, Roteb\"uhlstr.~87,
  70178 Stuttgart, Germany.}\\
{$^{\iETHZ}$Institute for Particle Physics, ETH Zurich,
  8093 Zurich, Switzerland.}\\
{$^{\iFR}$D\'epartement de Physique, Universit\'e de Fribourg,
  1700 Fribourg, Switzerland.}\\
{$^{\iTAIWAN}$Physics Department, National Tsing Hua University,
  Hsinchu 300, Taiwan.}\\
{$^{\iPRINCE}$Department of Chemistry, Princeton University, Princeton,
  NJ08544--1009, USA.}\\
\small{$^\dagger$ present address: Deutsches Zentrum f{\"u}r 
Luft- und Raumfahrt e.V. in der Helmholtz-Gemeinschaft,} \\[-0.5ex]
\small{70569 Stuttgart, Germany.} \\[-0.5ex]
\small{$^\spadesuit$ also at Ludwig-Maximilians-Universit\"at, Munich, Germany.}\\[-0.5ex]
\small{$^\S$ present address: LogrusData, Vienna, Austria.}\\[-0.5ex]
\small{$^\#$ also at Instituto Polit\'ecnico de Coimbra, ISEC, 3030--199,
  Portugal.}\\[-0.5ex]
\small{$^\diamondsuit$ present address: Forschungzentrum J\"ulich IKP-2 and 
RWTH Aachen University, Germany.}\\[-0.5ex]
\small{$^\odot$ present address: IAEA, Vienna, Austria.}\\[-0.5ex]
\small{$^\ddagger$ present address: Theiss Research Inc. La Jolla, CA, USA.}
}

\email{pohl@uni-mainz.de}

\recdate{Sept. 4, 2016}

\abst{Laser spectroscopy of the Lamb shift (2S-2P energy difference) in
light muonic atoms or ions, in which one negative muon \mumi\ is bound
to a nucleus, has been performed.
The measurements yield significantly improved values of the 
root-mean-square charge radii of the nuclei, owing to the large muon mass,
which results in a vastly increased muon wave function overlap with the nucleus.
The values of the proton and deuteron radii are 10 and 3 times more accurate
than the respective CODATA values, but 7 standard deviations smaller.
Data on muonic helium-3 and -4 ions is being analyzed and will give new insights.
In future, the (magnetic) Zemach radii of the proton and the helium-3 nuclei
will be determined from laser spectroscopy of the 1S hyperfine splittings,
and the Lamb shifts of muonic Li, Be and B
can be used to improve the respective charge radii.
}

\kword{laser spectroscopy, Lamb shift, muonic atoms, proton, deuteron, charge radius}

\begin{document}
\maketitle

\section{Introduction}

When negative muons (\mumi) are brought to rest in matter, they can form a
muonic atom or ion by ejecting all of the atom's electrons~\cite{Fiorentini:1978:Form2S}.
Thus, a single muon is bound to a nucleus.
The muonic atom/ion will quickly deexcite, mostly ending in the 1S ground state.
A few per cent of the muons can reach the metastable 2S state, however.
For light atoms, hydrogen to boron, the metastable 2S state can have 
a lifetime of microseconds to tens of nanoseconds~\cite{Mueller:1975:Quench2s,Arb:1984:2spop,Kirch:1997:muB}
that may make these metastable 2S atoms susceptible to laser spectroscopy~\cite{DiGiacomo:1969:MUP2S2P,Drake:1985:muLi_Be_B}.

For an isolated muonic atom/ion,
the 2S lifetime is limited by the muon lifetime of $2.2\,\mus$,
two-photon decay to the 1S ground state,
and nuclear muon capture, which is still small for light muonic ions~\cite{Measday:2001:NuclMuCap}.

In a gaseous environment like \Htwo, \Dtwo, He, etc., 
collisional effects may shorten the 2S lifetime~\cite{Carboni:1977:CollQuenchMup2S}.
Thus, our recent experiments in
muonic hydrogen~\cite{Pohl:2010:Nature_mup1,Antognini:2013:Science_mup2}, 
muonic deuterium~\cite{Pohl:2016:Science_mud} and 
muonic helium ions~\cite{Antognini:2011:Conf:PSAS2010,Nebel:2012:LEAP_muHe} were performed at gas pressures around 1\,hPa, where the 2S lifetimes are on the order of 1\,\mus~\cite{Pohl:2006:MupLL2S,Arb:1984:2spop}.

\begin{figure}
  %\centerline{
  %  \includegraphics[width=0.7\columnwidth]{LevelScheme_mup12_converted.eps}%
  %}
  \begin{minipage}[t]{.33\columnwidth}
    ~ (a) muonic hydrogen, \mup
  \end{minipage}
  \begin{minipage}[t]{.33\columnwidth}
    ~ (b) muonic deuterium, \mud
  \end{minipage}
  \begin{minipage}[t]{.33\columnwidth}
    ~ (c) muonic helium-4, \muHef
  \end{minipage}\\
  \begin{minipage}[t]{.33\columnwidth}
        \includegraphics[width=1.0\columnwidth]{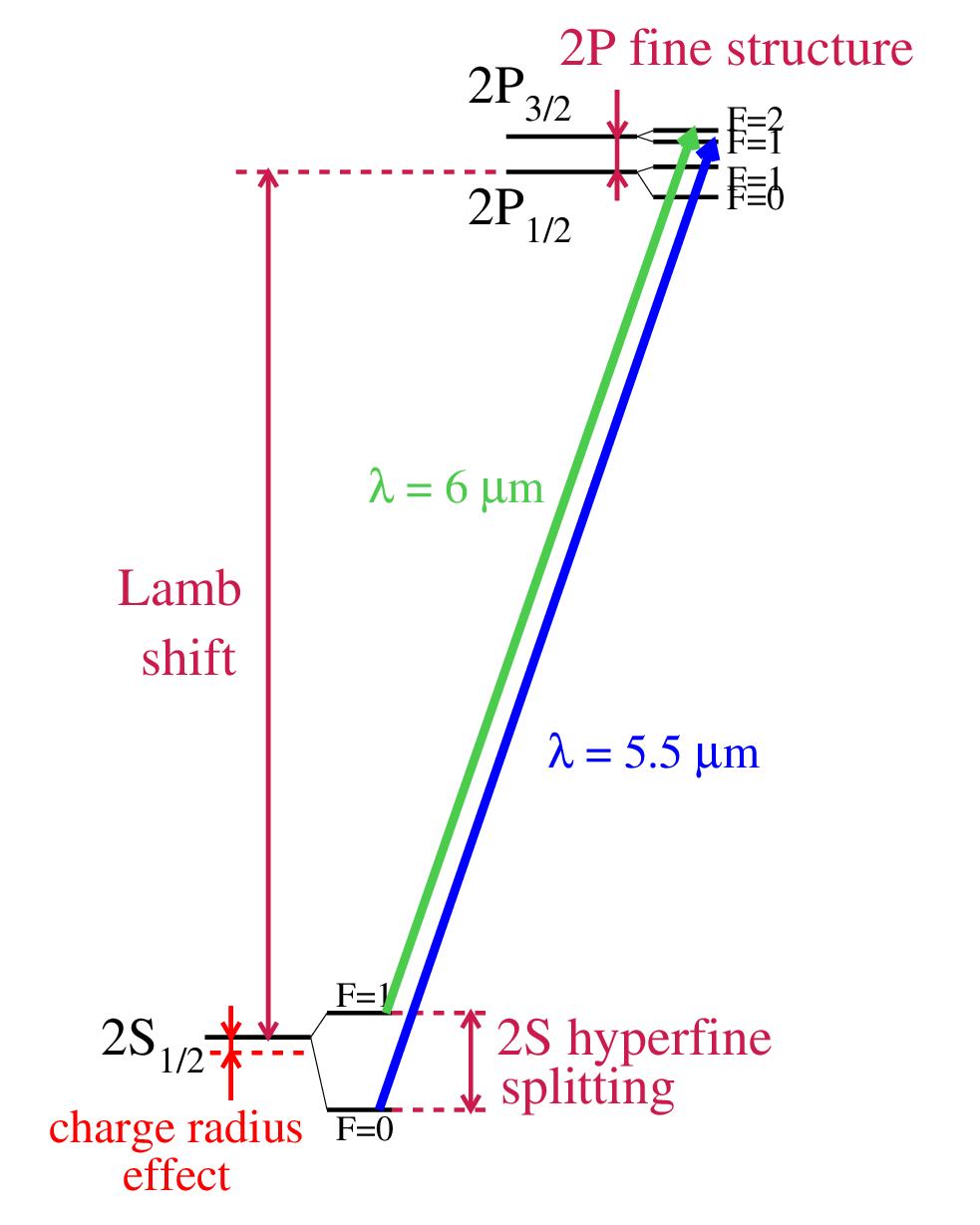}%
  \end{minipage}
  \begin{minipage}[t]{.33\columnwidth}
    \includegraphics[width=0.8\columnwidth]{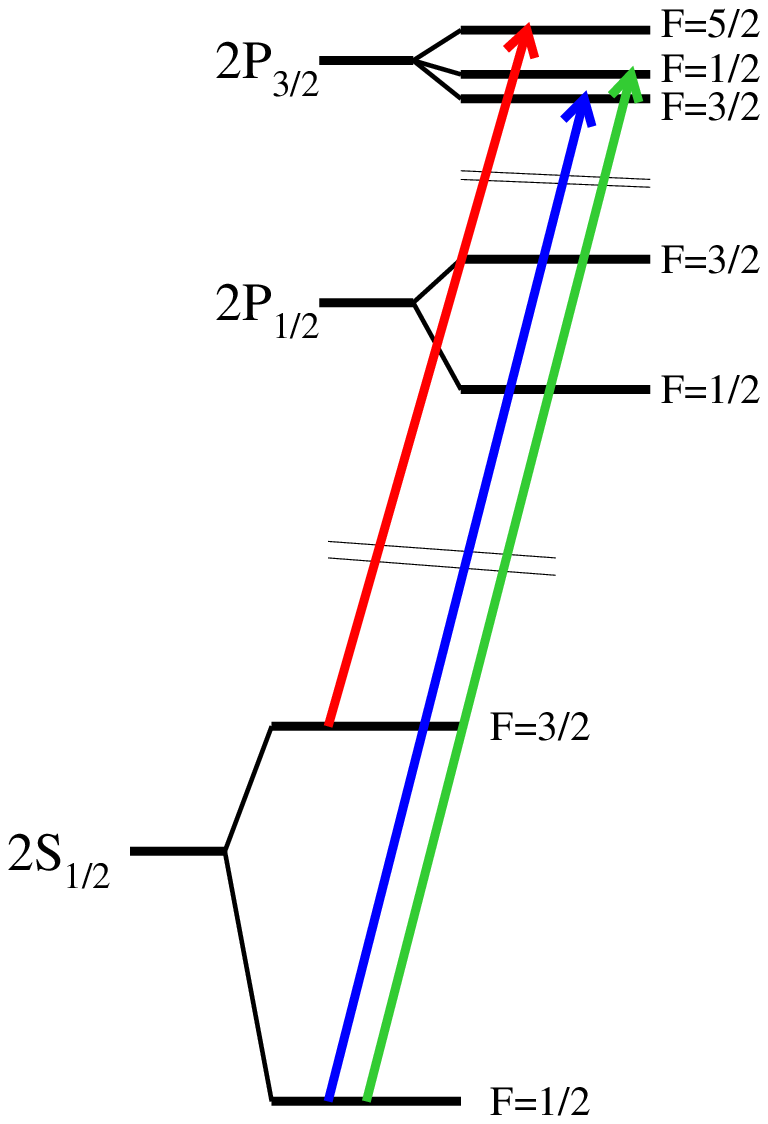}\\[-5ex]
    \setlength{\unitlength}{0.01\columnwidth}
    \begin{picture}(100,0)(0,0)
      %\put(0,0){\framebox(100,80){~}}
      \put(25,42){\fontsize{14pt}{14pt}\selectfont
        \color{red} $\nu_1$}
      \put(30,22){\fontsize{14pt}{14pt}\selectfont
        \color{Blue} $\nu_2$}
      \put(50,20){\fontsize{14pt}{14pt}\selectfont
        \color{ForestGreen} $\nu_3$}
    \end{picture}
  \end{minipage}
  \begin{minipage}[t]{.33\columnwidth}
    \includegraphics[width=0.83\columnwidth]{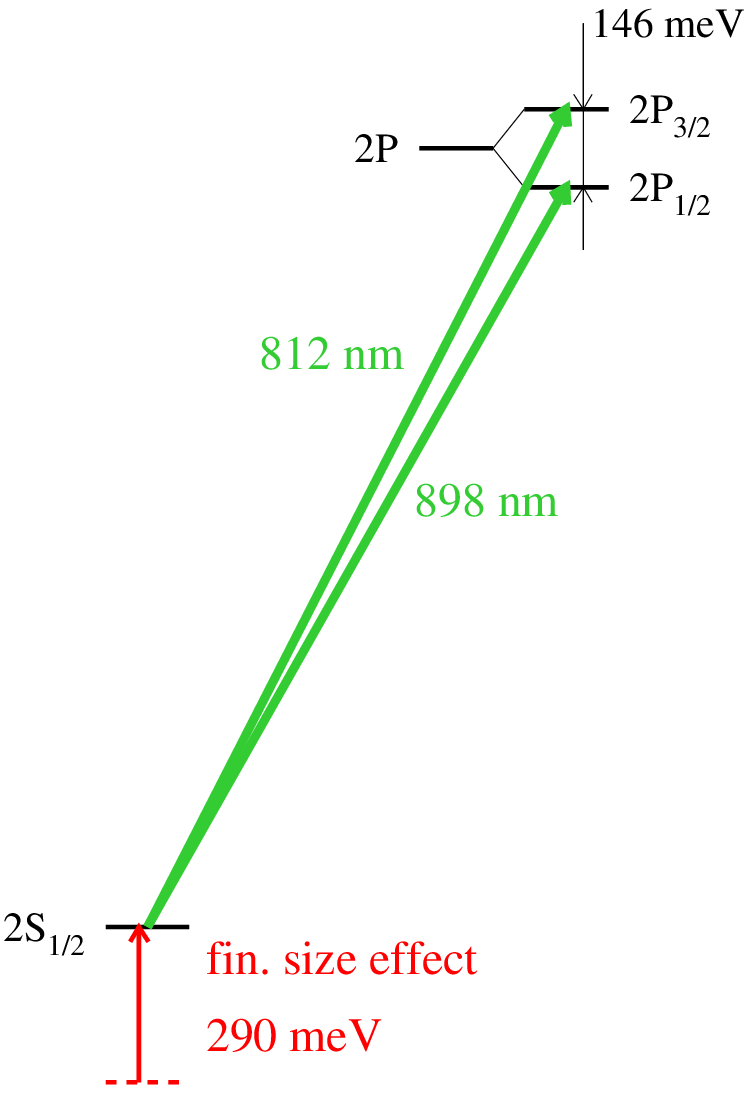}%
  \end{minipage}

  \caption{\label{fig:levels} Sketch of the $n=2$ levels in
    muonic hydrogen (left) and deuterium (center, not to scale) and
    muonic helium-4 ions (right).
    The nuclear charge radius shifts the 2S level upwards,
    as indicated for muonic hydrogen.
    The measured transitions are indicated.
    Muonic helium-3 ions, which were also measured, are not shown here.}
\end{figure}

The principle of the measurement is the to excite the
$2S\rightarrow 2P$ Lamb shift transition by a tunable pulsed laser system~\cite{Antognini:2005:6mumLaser,Antognini:2009:Disklaser}
and record the Lyman-$\alpha$ x-rays~\cite{Fernandes:2003:NIM,Ludhova:2005:LAAPDs,Fernandes:2007:JINST} from the
resulting $2P\rightarrow 1S$ deexcitation.

The laser frequency of the 2S-2P transition in light muonic atoms can then be
used to determine accurate values of nuclear charge radii or 
polarizabilities~\cite{Friar:1977:PRC16_1540,Friar:1978:Annals,BorieRinker:1982:muAtoms}.

\section{Lamb shift measurements and the charge radii}

\subsection{Proton radius from muonic hydrogen}
\begin{figure}[t]
  \begin{minipage}[t]{.5\columnwidth}
    \includegraphics[width=0.95\columnwidth]{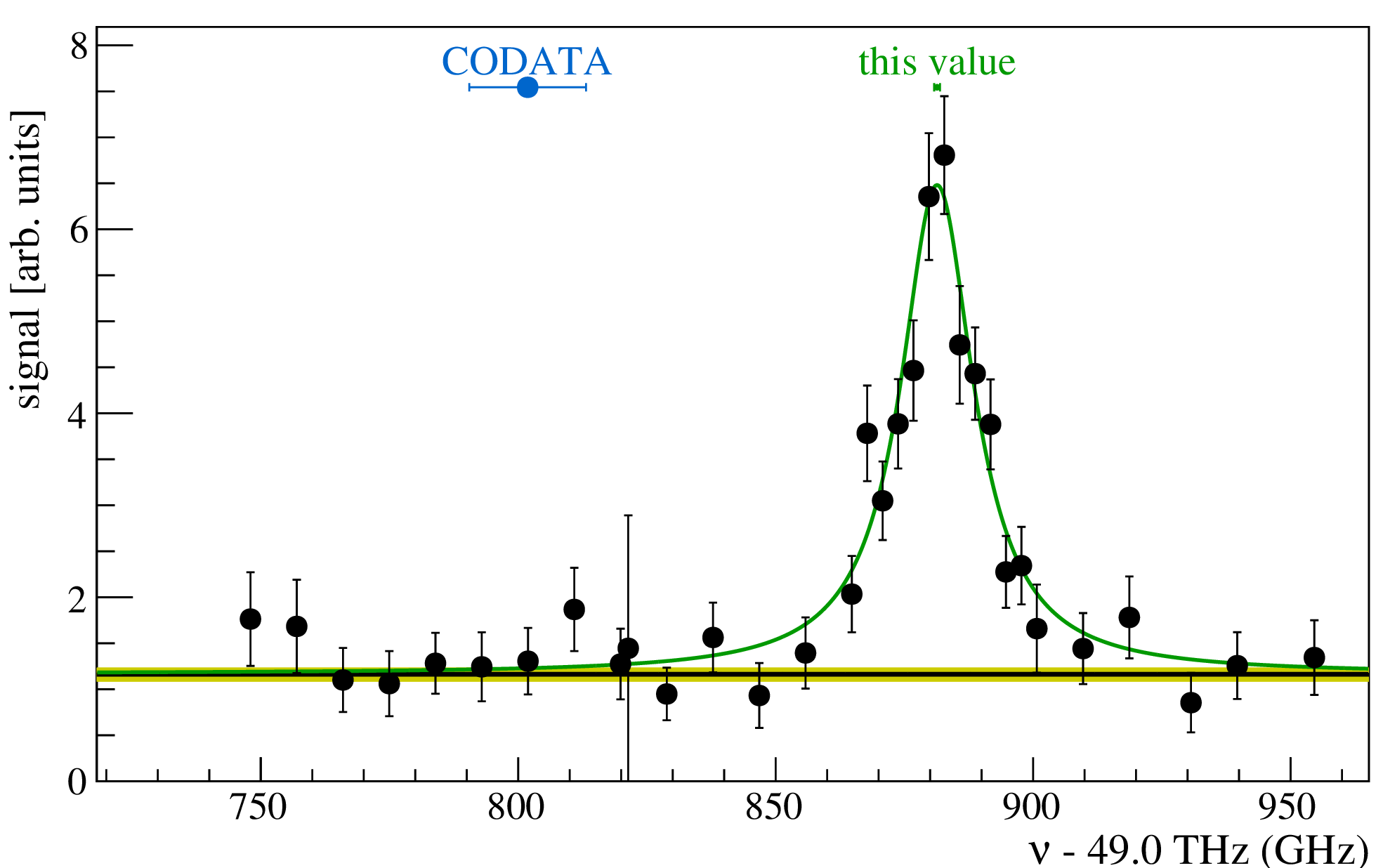}
  \end{minipage}
  \begin{minipage}[t]{.5\columnwidth}
    \includegraphics[width=0.95\columnwidth]{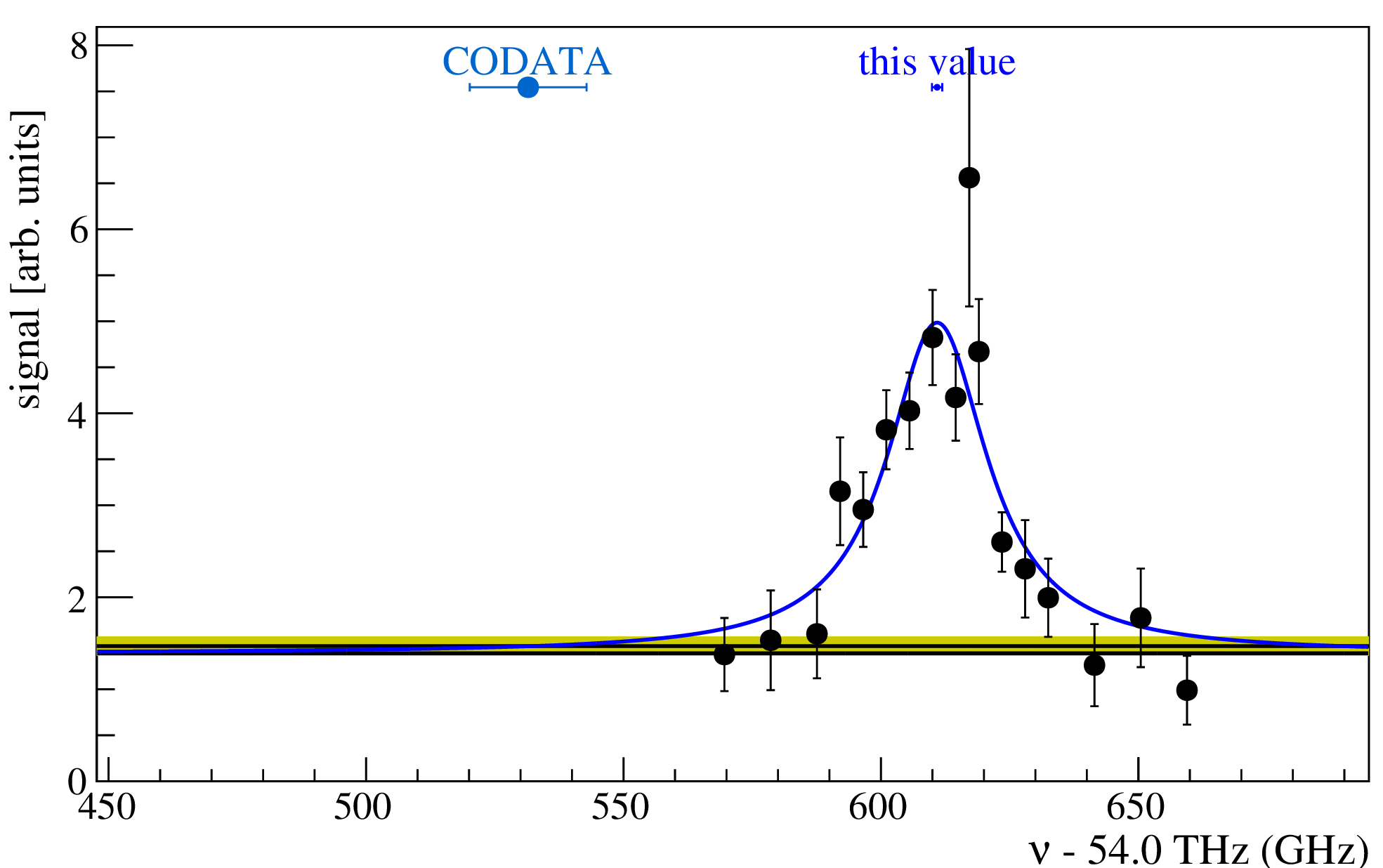}
  \end{minipage}
  \caption{\label{fig:res_muH}
    Two resonances measured in \mup\, see Fig.~\ref{fig:levels}(a).
    Left: $2S_{1/2}^{F{=}1}- 2P_{3/2}^{F{=}2}$ transition from the F=1 triplet
         state near $\lambda = 6.0$\,\mum\ ($49881.35 \pm 0.65$\,GHz).
    Right: The $2S_{1/2}^{F{=}0}- 2P_{3/2}^{F{=}1}$  transition from the 
         F=0 singlet state near $\lambda = 5.5$\,\mum\ 
         ($54611.16 \pm 1.03$\,GHz).
    The horizontal bar indicates the background level (with uncertainty), 
    including data taken without laser.
    The expected resonance positions calculated using the CODATA value of \rE(p)
    are 80\,GHz below the observed positions.
  }
\end{figure}

The Lamb shift in muonic hydrogen atoms, \mup, was first successfully measured 
at PSI in 2009.
A few unsuccessful searches were performed in the preceding years~\cite{Pohl:2005:MHL,Nebel:2007:SMH} at the wrong
laser wavelengths, because the expected resonance positions were based on a 
too large proton charge radius.
Finally, two transitions were measured~\cite{Pohl:2010:Nature_mup1,Antognini:2013:Science_mup2}.
Using the most recent theory from many authors~\cite{Pachucki:1996:LSmup,Pachucki:1999:ProtonMup,Borie:2005:LSmup,Martynenko:2005:mupHFS2S,Martynenko:2008:HFS_Pstates_mup,Eides:2001:PhysRep,Eides:2006:Book,Jentschura:2011:AnnPhys1,Jentschura:2011:relrecoil,Karshenboim:2012:PRA85_032509,Borie:2012:LS_revisited,Indelicato:2012:Non_pert,Karshenboim:2015:PhysRevD.91.073003}, summarized in Ref.~\cite{Antognini:2013:Annals}, we were able to deduce a value of the proton rms charge radius $\rE(p) = 0.84087(39)$~fm.
This value is $7\sigma$ smaller than the 2010-CODATA value~\cite{Mohr:2012:CODATA10}, see Fig.~\ref{fig:res_muH}.
This discrepancy is now known as the ``proton radius puzzle''~\cite{Pohl:2013:ARNPS,Carlson:2015:Puzzle}.

It is interesting to note that the muonic hydrogen value is smaller than both 
the value from spectroscopy of regular hydrogen alone~\cite{Pohl:2016:RdDeut}
and the widely accepted analyses of elastic electron-proton scattering data~\cite{Bernauer:2010:NewMainz,Sick:2003:RP,Sick:2014:Tail,Sick:2014:PRC,ArringtonSick:2015:JPCRD}. However, several analyses of e-p scattering data have found radii that are in agreement with the muonic hydrogen value~\cite{Belushkin:2006qa,Lorenz:2012:Closing_in,Lorenz:2015:Constaints_FF,Griffioen:2016:SmallRp,Higinbotham:2016:Rp_ep,Horbatsch:2016:ep_scatt}.

\begin{figure}[t]
  \begin{minipage}[t]{.5\columnwidth}
    \includegraphics[width=0.95\columnwidth]{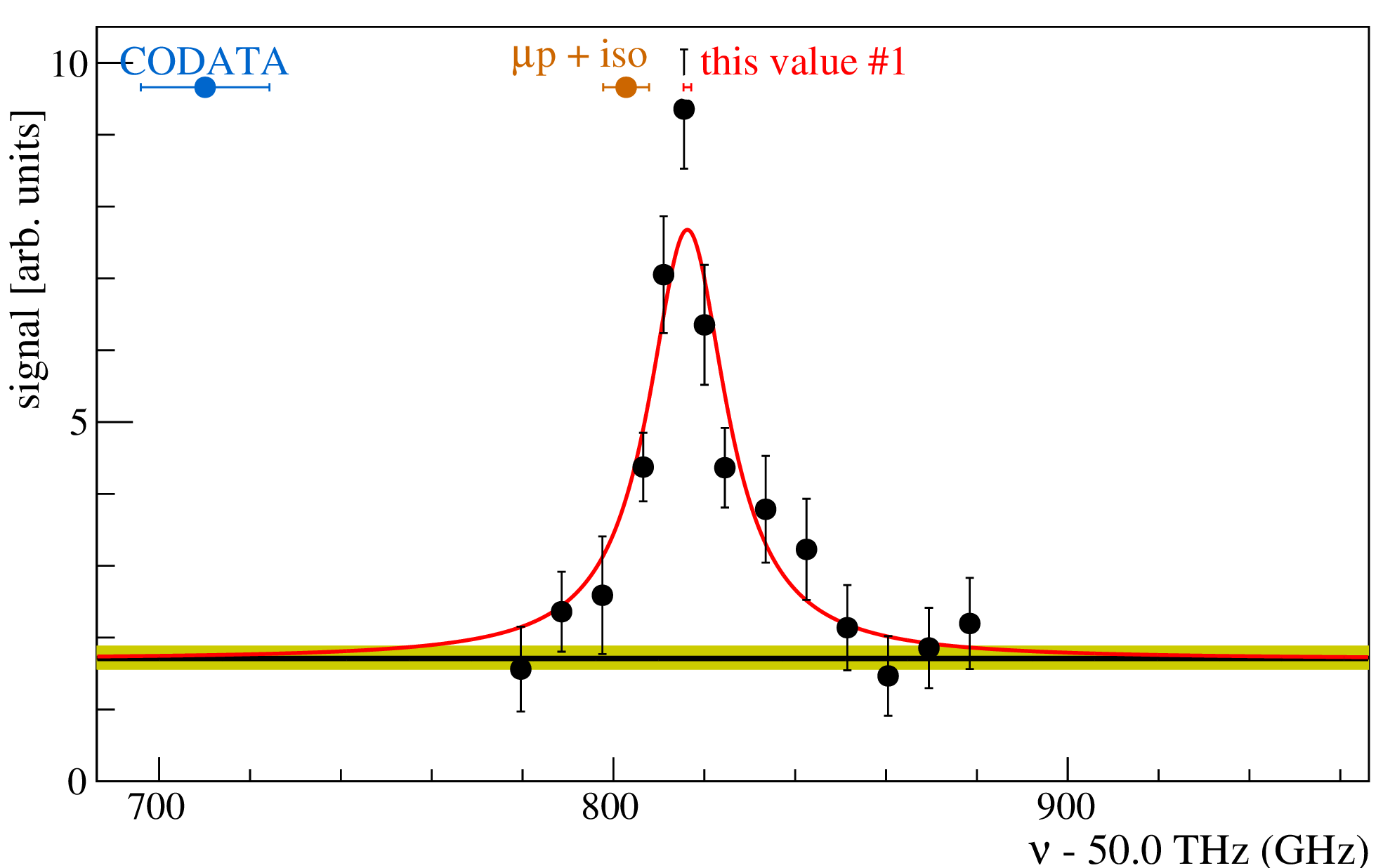}
  \end{minipage}
  \begin{minipage}[t]{.5\columnwidth}
    \includegraphics[width=0.95\columnwidth]{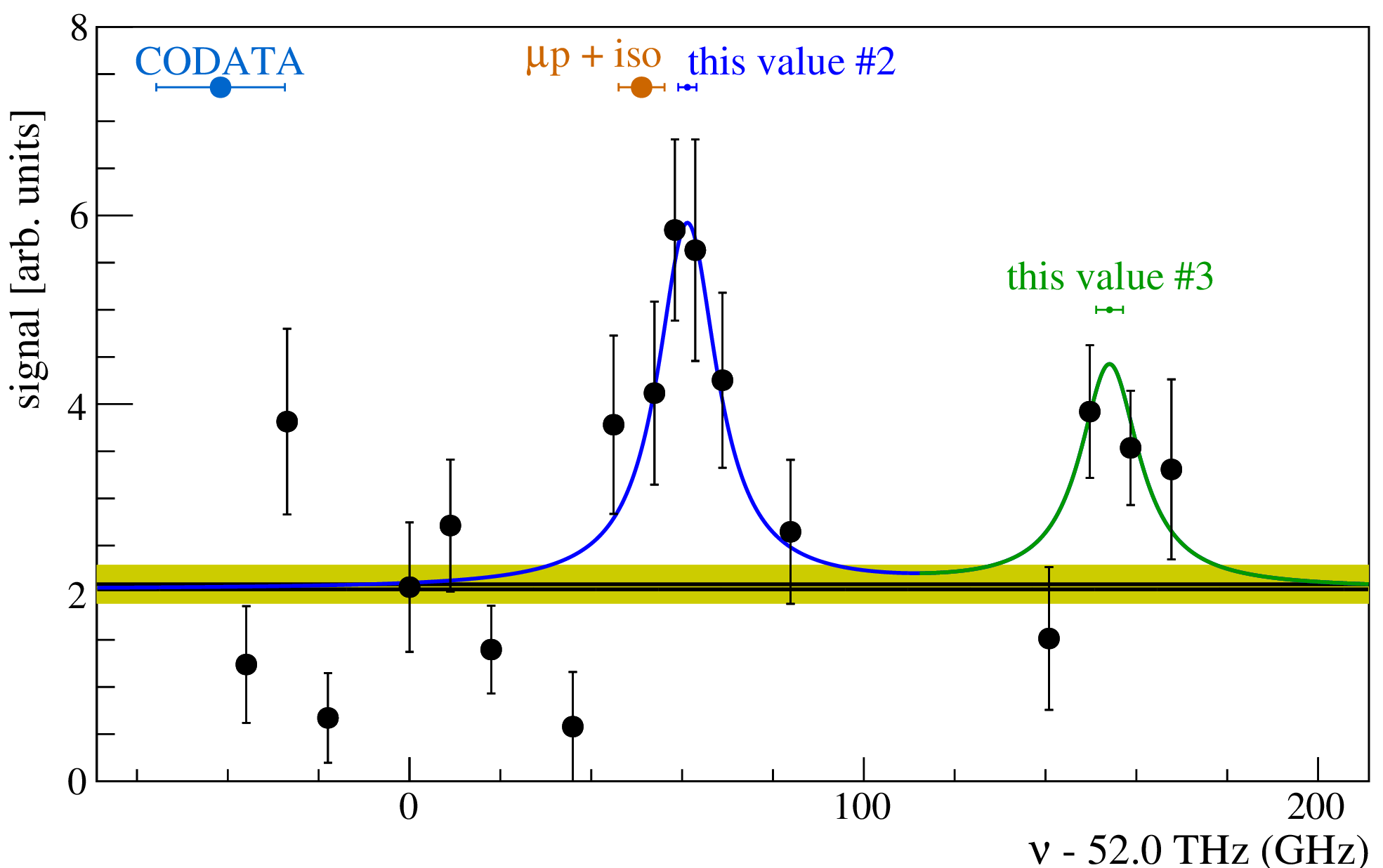}
  \end{minipage}
  \caption{\label{fig:res_muD}
    Three resonances measured in \mud\, see Fig.~\ref{fig:levels}(b).
    Left: \DLineOne, and right: \DLineTwo, and \DLineThree.
    The colors are the same as in Fig.~\ref{fig:levels}(b).
    The horizontal bar indicates the background level (with uncertainty), 
    including data taken without laser.
    The expected resonance positions calculated using the CODATA value of \rE(d)
    are 100\,GHz below the observed positions.
    In contrast, the {\em proton} radius \rE(p) from muonic hydrogen, 
    when combined with the electronic isotope shift,
    yields a prediction of the resonance position ``\mup\ + iso'' which is
    within $2.6\sigma$ of the observed ones.
    The muonic proton and deuteron radii are hence compatible.
  }
\end{figure}
  
\subsection{Deuteron radius from muonic deuterium}
Very recently, theory calculations in muonic deuterium~\cite{Borie:2005:LSmud,Krutov:2011:PRA84_052514,Borie:2012:LS_revisited,Faustov:2014:HFS_mud,Faustov:2015:HFS_2P_mud}, and in particular the difficult deuteron polarizability effects~\cite{Pachucki:2011:PRL106_193007,Friar:2013:PRC88_034004,Hernandez:2014:PLB736_344,Carlson:2014:PRA89_022504,Pachucki:2015:PRA91_040503}
became accurate enough to determine a deuteron charge radius from our
measurements in muonic deuterium.
Again, we summarized these earlier works~\cite{Krauth:2016:Annals}.
The $2S$ Lamb shift in muonic deuterium is
\begin{equation}
  \label{eq:LStheo}
  \Delta E_\mathrm{LS} = 228.7766(10)\,\mathrm{meV} + \Delta E^\mathrm{TPE}_\mathrm{LS} - 6.1103(3)\, \rd^2\,\mathrm{meV/fm^2}.
\end{equation}

\begin{figure}
  \begin{minipage}[t]{.5\columnwidth}
    \includegraphics[width=0.95\columnwidth]{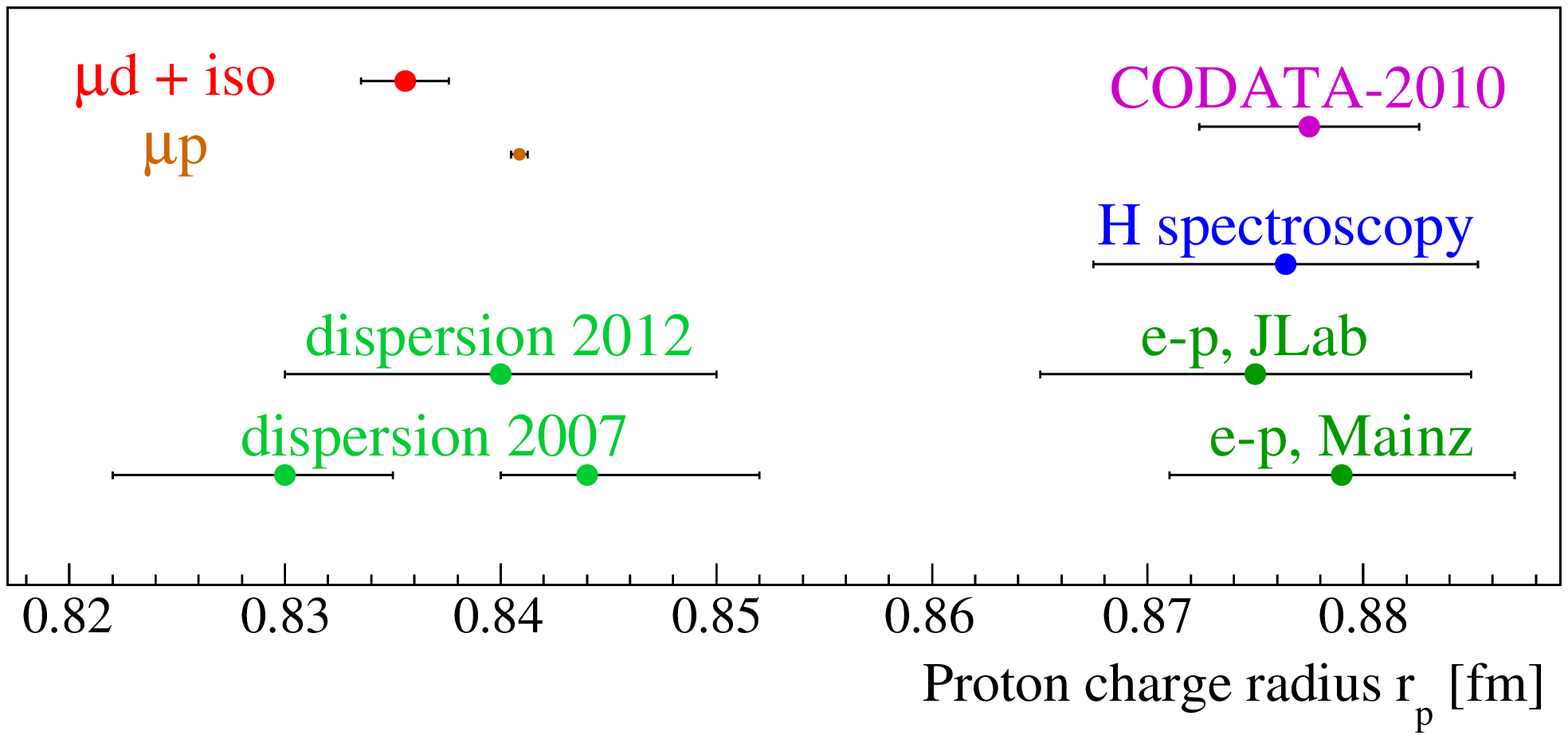}
  \end{minipage}
  \begin{minipage}[t]{.5\columnwidth}
    \includegraphics[width=0.95\columnwidth]{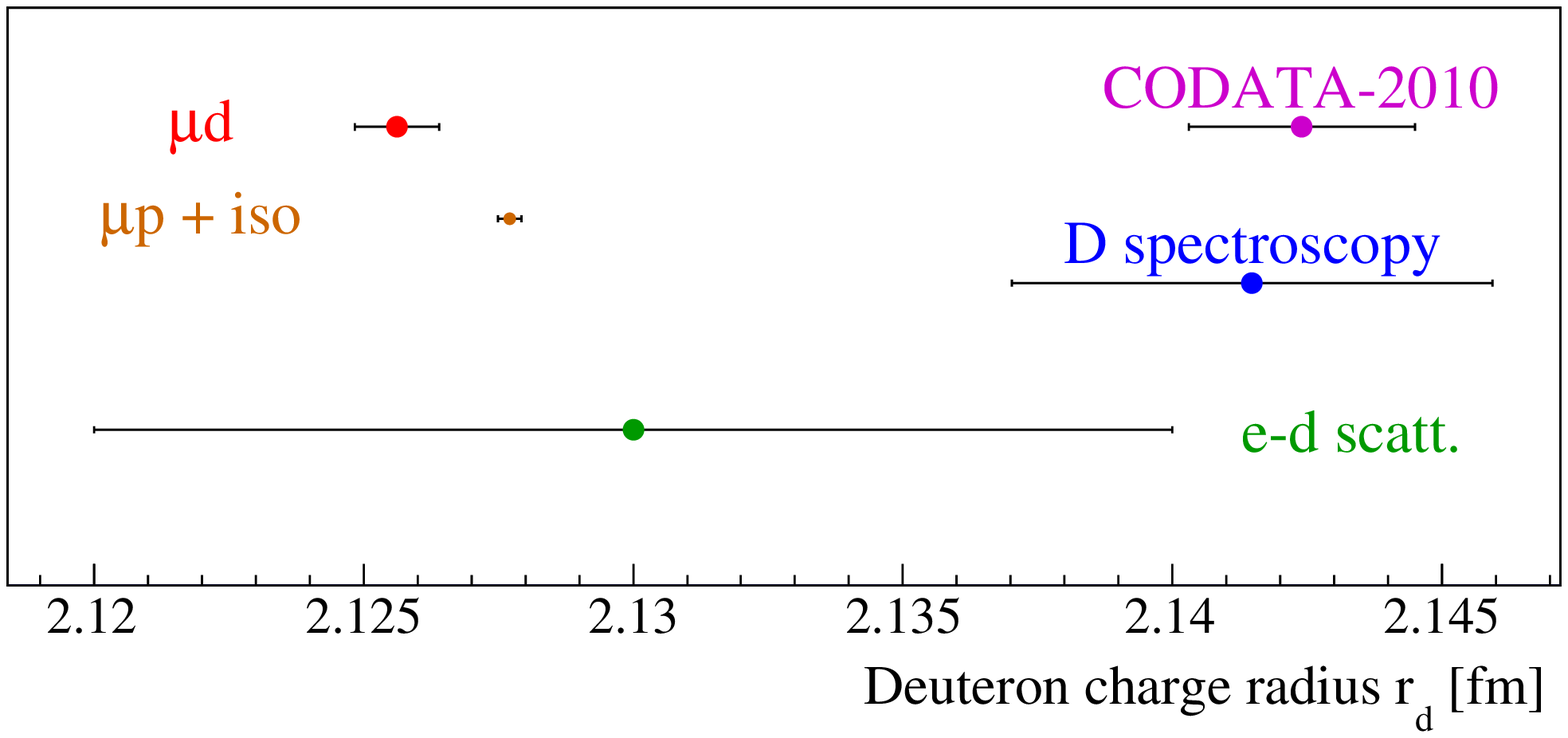}
  \end{minipage}
  \caption{\label{fig:radii}
    Root-mean-square charge radii of the proton (left) and deuteron (right)
    from various sources. 
    The muonic values are within $2.6\sigma$ compatible with each other, when
    considering the 1S-2S isotope shift in regular hydrogen and deuterium~\cite{Parthey:2010:PRL_IsoShift,Jentschura:2011:IsoShift}.
    Values ``H spectroscopy'' and ``D spectroscopy'' are uncorrelated, as
    explained in~\cite{Pohl:2016:RdDeut}.
    The discrepancy between muonic and electronic deuterium is hence a new
    discrepancy.
  }
\end{figure}

Using the calculated contribution from two-photon exchange, also known as the 
deuteron polarizability contribution
\begin{equation}
  \label{eq:TPE_theo}
  \Delta E^\mathrm{TPE}_\mathrm{LS}\mathrm{(theo)} = 1.7096(200)\,\mathrm{meV}
\end{equation}
we determine the deuteron rms charge radius from a measurement of three 
2S-2P transitions in muonic deuterium~\cite{Pohl:2016:Science_mud}
\begin{equation}
  \label{eq:Rd_muD}
  \rd~[ \mud ] = 2.12562 (13)_\mathrm{exp} (77)_\mathrm{theo} \,\mathrm{fm}.
\end{equation}
Similar to the proton radius puzzle, this value from muonic deuterium is
$7\sigma$ smaller than the CODATA-2010 deuteron charge radius,
see Fig.~\ref{fig:radii}.

We can also determine a value of the deuteron charge radius
\rd~[\mup~+~iso] = 2.12771\,(22)\,fm
 from combining the
{\em electronic} isotope shift of the 1S-2S transition 
in H and D~\cite{Parthey:2010:PRL_IsoShift,Jentschura:2011:IsoShift},
\begin{equation}
  \label{eq:HDiso}
  \rd^2 - \rp^2 = 3.82007(65)\,\mathrm{fm}^2
\end{equation}
and the proton charge radius from muonic hydrogen~\cite{Antognini:2013:Science_mup2}. Using this value in Eq.~(\ref{eq:LStheo}), and the measured Lamb shift in muonic deuterium, we can determine the deuteron polarizability from experiment
\begin{equation}
  \label{eq:TPE_exp}
  \Delta E^\mathrm{TPE}_\mathrm{LS}\mathrm{(exp)} = 1.7638(68)\, \mathrm{meV}
\end{equation}
which is three times more accurate than the
theoretical value~\cite{Krauth:2016:Annals},
and thus a benchmark for future high-precision calculations.

\subsection{Muonic helium ions}

\begin{figure}
  % trim = left, bottom, right, top
  %\includegraphics[width=0.95\columnwidth, trim= 250 280 30 0, clip=true]{muHe1_b.eps}
  \centerline{
  \includegraphics[width=0.5\columnwidth, trim= 250 280 30 0, clip=true]{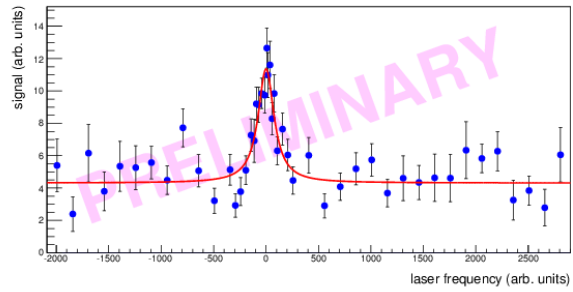}
  }
  \caption{\label{fig:res:mu4He1}
    Preliminary resonance plot of the 1st muonic $^4$He resonance (online data).
    The alpha particle charge radius extracted from this resonance will be 
    about five times more accurate than the one from elastic electron
    scattering.
    The muonic charge radius uncertainty is completely dominated by the 
    accuracy of the nuclear polarizability.
  }
\end{figure}

In two beam times in 2013 and 2014 we have measured several transition frequencies in
muonic helium-3 and -4
ions~\cite{Antognini:2011:Conf:PSAS2010,Nebel:2012:LEAP_muHe}.
An example of a measured resonance is given in Fig.~\ref{fig:res:mu4He1}.
Again, by comparing the theoretical prediction of many
authors~\cite{Karshenboim:2010:NRalpha5_mup,Karshenboim:2010:JETP_LBL,Borie:2012:LS_revisited,Krutov:2014:JETP120_73,Karshenboim:2012:PRA85_032509,Korzinin:2013:PRD88_125019,Jentschura:2011:PRA84_012505,Jentschura:2011:relrecoil,Jentschura:2011:SemiAnalytic,Elekina:2011:FSmu4He,Ji:2013:PRL111},
as summarized by us~\cite{Diepold:2016:TheoMu4He}, values
of the alpha-particle and helion charge radii can be deduced with an accuracy
of about $3\times 10^{-4}$. These will be compared with the values from
electron scattering~\cite{Sick:2008:rad_4He,Sick:2015:JPCRD}
and He spectroscopy~\cite{Rooij:2011:HeSpectroscopy,CancioPastor:2012:3He-4He,Shiner:1995:3He,Pachucki:2012:3He,Herrmann:2009:He1S2S,Kandula:2011:XUV_He}.

Besides providing important information useful for the resolution of the
proton radius puzzle, these radii are interesting parameters to be
compared with scattering results~\cite{Sick:2008:rad_4He,Sick:2015:JPCRD},
few-nucleon ab-initio calculations~\cite{Epelbaum:2009:NuclTheo},
and effective nuclear
theories~\cite{Ji:2013:PRL111,Lu:2013:RMP,Ji:2014:FewBodySyst,Hagen:2016:48Ca}.
Moreover, the muonic radii open the way to enhanced bound-state QED tests for
one- and two-electron systems from measurements in ``regular'' He and
He$^+$ ions~\cite{Herrmann:2009:He1S2S,Kandula:2011:XUV_He,Rooij:2011:HeSpectroscopy,CancioPastor:2012:3He-4He,Shiner:1995:3He,Pachucki:2012:3He,Antognini:2011:Conf:PSAS2010,Jentschura:2005:NRQED}.
They will also be used to disentangle the 4$\sigma$ discrepancy
between several $^3$He$-^4$He isotopic shifts measurements~\cite{Rooij:2011:HeSpectroscopy,CancioPastor:2012:3He-4He,Shiner:1995:3He,Pachucki:2012:3He}, 
and serve as anchor points to provide absolute radii for $^6$He and
$^8$He halo nuclei when combined with the corresponding isotopic
shifts measurements~\cite{Lu:2013:RMP}.
%% %

\section{Hyperfine splitting of muonic hydrogen and helium}

\begin{figure}
  \centerline{
  \includegraphics[width=0.5\columnwidth]{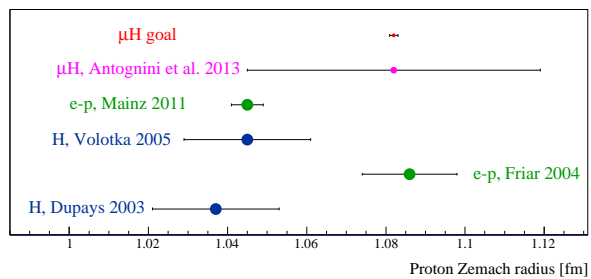}
  }
  \caption{\label{fig:Rz}
    Proton Zemach radius determinations.
    Accurate values exist from the 1S HFS in hydrogen (21~cm line) and
    elastic electron-proton scattering.
    The first result from muonic hydrogen will be improved in upcoming
    experiments.
  }
\end{figure}

Previously we presented the determination of the rms
charge radii from the measurement of 2S-2P Lamb shift transitions in muonic
atoms.
In a similar way, spectroscopy of the hyperfine splitting (HFS) in
muonic and ``regular'' atoms can be used to deduce precise values of
the Zemach radii (of various nuclei).
The Zemach radius $R_Z$ being defined as an integral of the charge and
magnetic form factors ($G_E$, $G_M$)
\begin{equation}
R_Z=-\frac{4}{\pi}\int_0^\infty \frac{dQ}{Q^2} \; \Big( G_E(Q^2)
\frac{G_M(Q^2)}{1+\kappa_p}-1 \Big) \; ,
\label{eq:def-Rz-1}
\end{equation}
where $\kappa_p$ is the proton anomalous magnetic moment and $Q$ the
momentum exchange, contains information about the magnetization
distribution inside the nucleus.
In fact, in a non-relativistic approximation $R_Z$ can be expressed, by
the convolution between nuclear charge and magnetic distributions
$\rho_E(r)$ and $\rho_M(r)$
\begin{equation}
R_Z=\int d^3 {\bf r} \; |{\bf r}| \int d^3{\bf r}' \rho_E(\mathbf{r}-\mathbf{r'}) \rho_M(\mathbf{r'}).
\label{eq:def-Rz-2}
\end{equation}

A first, rather crude measurement of the hyperfine splitting in muonic
hydrogen has been achieved by the CREMA collaboration in
2013~\cite{Antognini:2013:Science_mup2}, see Fig.~\ref{fig:Rz}.
Here, the difference of two 2S-2P Lamb shift transitions was used
to determine the 2S-HFS with an accuracy of $\sim 220$~ppm.

As a next step, the CREMA collaboration is aiming at the measurements
of the ground-state HFS in \mup{} and \muHet{} with relative accuracy
of about 1~ppm.
Two other groups, are also planning to measure the 1S-HFS in \mup{} so
that, differently to the 2S-2P case, comparison between muonic results
will become possible.
Having different experimental schemes, these three experiments allow
three independent determinations of the Zemach radius.
In the following we depict the three methods proposed by the three
collaborations to measure the ground-state HFS.

\subsection{CREMA: Diffusion method}
\begin{figure}
  \begin{minipage}[t]{.4\columnwidth}
    \includegraphics[width=0.9\columnwidth]{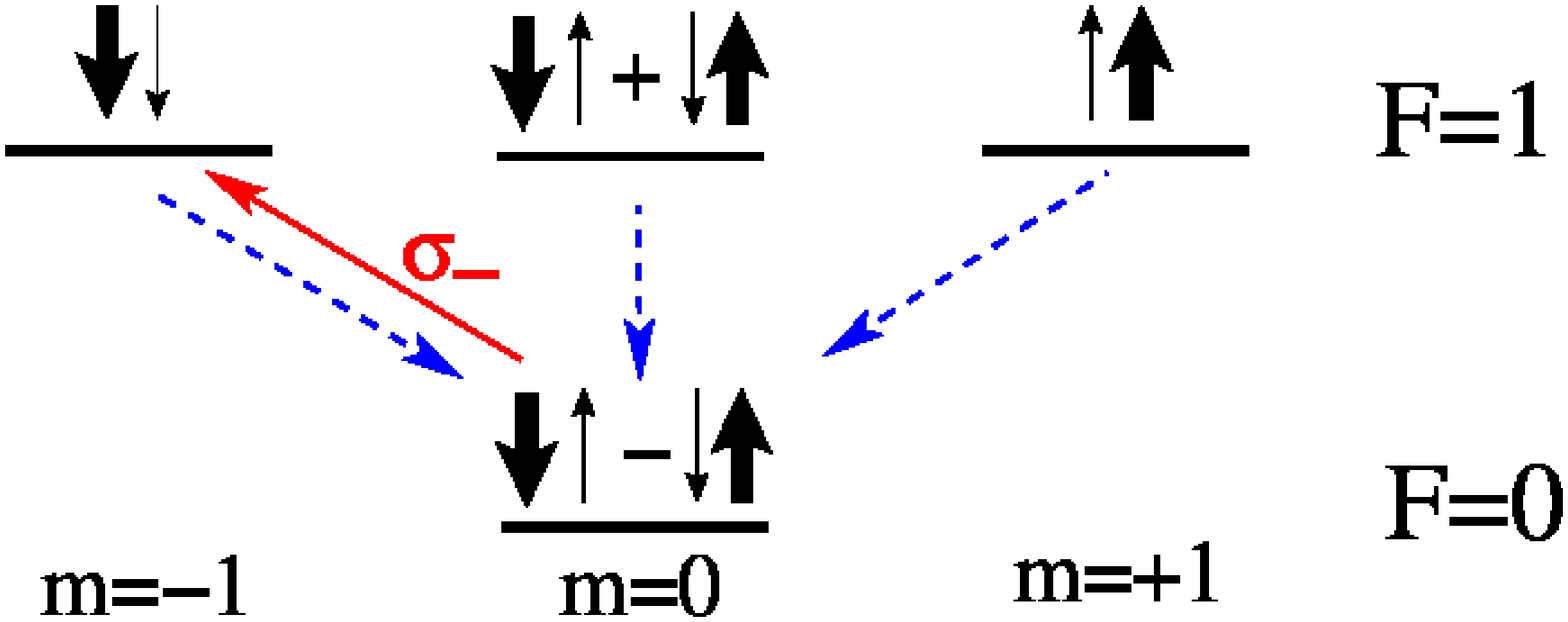}
  \end{minipage}
  \begin{minipage}[t]{.6\columnwidth}
    \includegraphics[width=0.9\columnwidth]{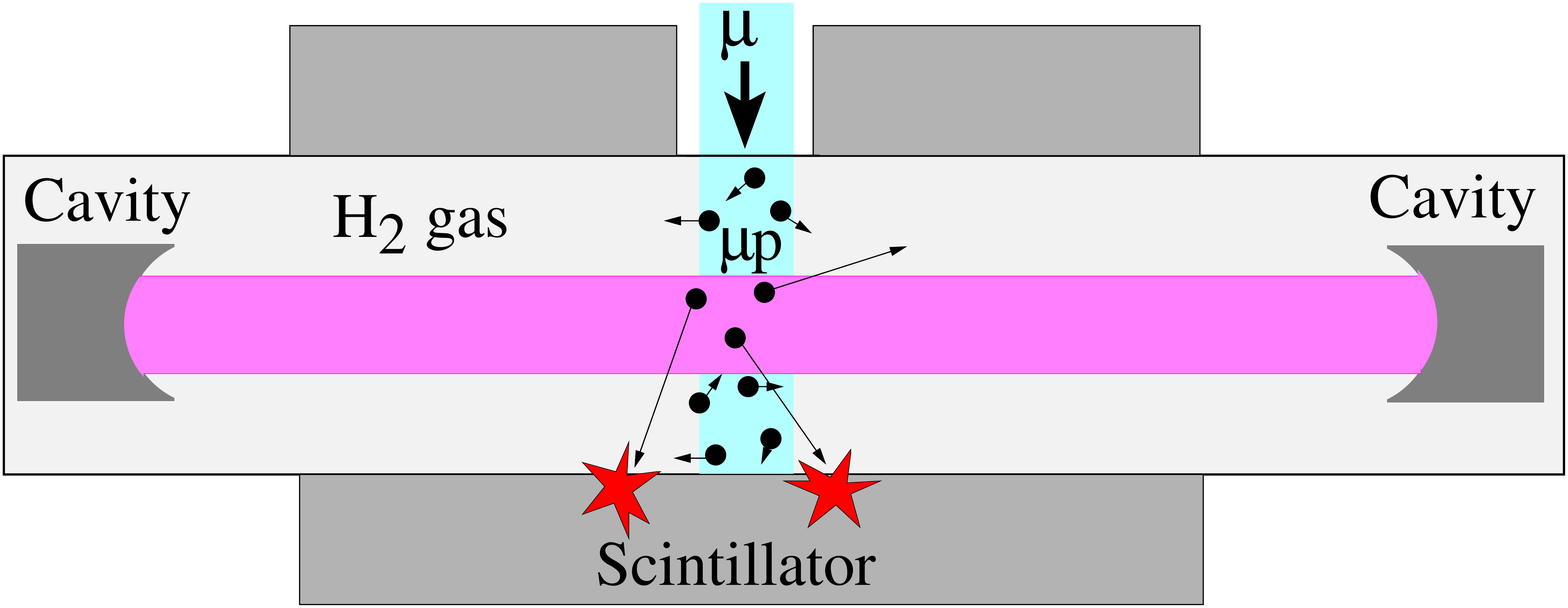}
  \end{minipage}
  \caption{\label{fig:HFS-1} (Left) Level scheme of the \mup\ ground
    state divided into triplet ($F=1$) and singlet ($F=0$)
    sublevels. The spin composition of the sub-states is also
    given. The red solid-line arrows represent the transitions driven
    by the polarized laser light, the dashed blue arrows the quenching
    of the triplet states into the singlet state caused by collision
    of the \mup\ atoms with the H$_2$ gas.  (Right) Principle and
    setup (not to scale) of the 1S-HFS experiment of the CREMA
    collaboration.  Negative muons stop in the few~mm long gas target
    at 50~K and 500~mbar.  The \mup{} atoms (black dots) excited by
    the laser light (given in pink) have larger kinetic energy and
    diffuse from the illuminated volume to the target walls coated
    with high-Z material. The thermalized \mup{} atoms are much
    slower.  Scintillators are used to observe the cascade events
    after formation of the excited $(\mu Z)^*$ atoms.  }
\end{figure}
The CREMA collaboration experiment is performed at the proton
accelerator facility of the Paul Scherrer Institute, Switzerland.
A continuous negative muon beam of~10 MeV/c momentum is stopped in a
hydrogen (H$_2$) gas target at cryogenic temperatures (50~K, 500 mbar).
The $\mu^-$ stopping in the H$_2$ gas forms \mup{} atoms in highly
excited states ($n\sim14$).
After the prompt muonic cascade, the 1S-state is populated with
a~statistically distributed occupancy of the four sub-levels (triplet
and singlet) as visible in Fig.~\ref{fig:HFS-1} (left).
Shortly afterwards, the higher lying triplet states are collisionally
quenched to the singlet state (with a rate of 25/\mus{} at 1~bar and
273~K)~\cite{Cohen:1991:mupXsect2}),
so that at the laser pulse arrival time the
\mup{} atom is basically in the singlet ground-state.
At 1.5~\mus{} after the \mup{} formation, the laser system delivers a
pulse of 1~mJ energy at 6.7~\mum{} wavelength to drive the hyperfine
splitting transition from $F=0$ to $F=1$.
Shortly afterwards (within about 100~ns), the higher lying triplet
state is collisionally quenched back to the singlet state.
In this process, the HFS transition energy is converted into kinetic
energy: the \mup{} atom gain a kinetic energy of about 120~meV.
For a sufficiently small target, as shown in Fig.~\ref{fig:HFS-1}
(right) the faster moving \mup{} starting from the laser illuminated
region may reach the target walls coated with high-Z material before
$\mu^-$ decay occurs~\cite{Abbott:1997:mupDrift}.
Subsequent $\mu^-$ transfer from the \mup{} to a high-Z atom occurs,
leading to the formation of a $(\mu Z)^*$ atom followed by its 
de-excitation producing MeV X-rays.
Therefore if the laser frequency is resonant with the HFS transition
MeV X-rays will be detected right after the laser pulse has
illuminated the muon stop distribution.
A resonance curve is attained by plotting the number of the cascade
signals (from the MeV X-rays) induced in the surrounding scintillators
in a short time window ($\sim300$~ns) after the laser excitation
versus the laser frequency.

\subsection{FAMU: Energy dependent muon transfer method}

The methodology of the FAMU collaboration~\cite{Bakalov:2015:ExpHFS, Bakalov:2015:Xfer} differs from the one of the
CREMA collaboration mainly in the detection scheme used to expose the
laser induced occurrence of the hyperfine transition.

In this experiment which is performed at the RIKEN-RAL muon facility
of the Rutherford Laboratory, UK, a pulsed muon beam is stopped in a
hydrogen target with 0.045 LHD (liquid hydrogen density).
Shortly after its formation the muonic atom ends up in the singlet
sub-level of the ground state due to the muonic cascade and the
collisional quenching from the triplet to the singlet ground states.
A laser pulse is then used to drive the HFS transition, i.e. to excite
the muonic atom from the $F=0$ to the $F=1$ state.
The atoms excited by the laser in the triplet state, de-excite back
to the $F = 0$ state through collisions with the H$_2$ molecules
gaining in this process about 120~meV of kinetic energy.

The key difference between the CREMA schema and the FAMU scheme is in
the method used to distinguish the two classes of atoms: the one
which have been laser excited by the laser light, from the one which
thermalize at the hydrogen gas temperature.
Instead of using the diffusion method of CREMA, the FAMU collaboration utilizes
the energy dependence of the muon transfer from the hydrogen atom to a
higher-Z atom which is mixed with a few \% concentration in the H$_2$
gas.
The muon transfer rate from the \mup\ to some higher-Z gases (CO$_2$,
Ar...) has recently been investigated in a beamtime at the RIKEN-RAL
facility~\cite{Adamczak:2016:transfer:measurement}.
Data analysis is ongoing.

The higher-Z muonic atoms  which are formed in highly excited
states de-excite down to the ground state producing X-rays in the
10~keV-1~MeV energy regime that can be efficiently detected with a
scintillator or Germanium system.
The resonance curve is therefore obtained by plotting the number of
X-rays from the ``contaminant'' gas versus the laser frequency.

\subsection{J-PARC: Decay asymmetry method}

The J-PARC~-~RIKEN collaboration is planning to use the electron
from the muon decay to expose the occurrence of the laser induced
transition~\cite{Sato:2015:ExpHFS,Saito:2012:HFS:Laser}.
As mentioned before, after its formation the \mup\ atoms end up quickly
in the ground state populating both triplet and singlet states.
In this experiment the H$_2$ gas target density is reduced to
0.0001~LHD to minimize the collisional quenching from the triplet
into the singlet state.
At this density the lifetime of the triplet state is of about 500~ns.
The circular polarized laser light which illuminates the target
1~\mus{} after the muonic atom formation is inducing a transition from
the singlet to a triplet state.
This transition which give rise to a muon spin flip, results in a
change of the average muon polarization (which is zero prior to the
laser pulse).
This change of the muon polarization after the laser transition can be
exposed by detecting the electron from muon decay.
In fact, due to the parity violation in the muon decay, the electron
from muon decay is emitted preferentially anti-parallel to the muon
spin.
Therefore the change of the muon polarization induced by the laser
light results in a change of the electron distribution 
which can be observed by using electron detectors.
A resonance curve can be obtained by plotting the so called asymmetry
(difference in counts between two detectors) as a function of the
laser frequency. The change of the electron distribution will be visible in this asymmetry plot.

The choice of the low gas density results from a trade-off between
muon stopping volume and efficiency, and the collisional quenching rate from
the polarized triplet (after the laser excitation) into the singlet
(unpolarized). Therefore the time window where the electrons have to be detected
is only of about 1~\mus{} given the above-mentioned depolarization rate.

\subsection{Impact of the HFS measurements}

By comparing the measured HFS in \mup\ and \muHet\ with the
corresponding theoretical predictions~\cite{Eides:2006:Book, Karshenboim:2005:PPS, Martynenko:2008:HFS:He} the two-photon-exchange TPE
contribution with a relative accuracy of $10^{-4}$ can be extracted.
From the TPE the polarizability contribution can be extracted if a
Zemach radius is assumed from
electron scattering~\cite{Distler:2011:Zemach, Sick:2014:Zemachmoments}
or from regular atom spectroscopy~\cite{Dupays:2003:Zemach,Volotka:2005:Zemach}.
Alternatively the Zemach radii can be extracted if the
polarizabilities are assumed from theory.
A comparison of these radii and polarizability contributions with
theory 
(Chiral perturbation theory~\cite{Hagelstein:2016},
dispersion relations~\cite{Carlson:2008:p_struct_HFS, CarlsonNazaryan:2011},
lattice calculations~\cite{Alexandrou:2016:lattice},
and ab-initio few nucleon theories~\cite{Ji:2014:FewBodySyst})
offers the possibility to improve our knowledge of the
low energy structure of the of the proton and of one of the 
simplest nuclei, the helion.

\section{Conclusions}

Laser spectroscopy of muonic atoms can be used for the determination
of nuclear parameter with high accuracy, to be compared with the 
values from electron scattering and regular atom spectroscopy.
A large activity both, theoretical and experimental,
have been triggered by the observed discrepancy, which remains still
unsolved.
Many new activities are underway, such as
spectroscopy in (regular) atomic hydrogen~\cite{Vutha:2012:H2S2P,Beyer:2013:AdP_2S4P,Beyer:2013:Conf:ICOLS,Galtier:2015:JPCRD,Yost:2016:1S3S}
and molecular hydrogen (and its isotopes)~\cite{Schiller:2014:MolClock,Dickenson:2013:H2vib,Biesheuvel:2015:HDplus},
new elastic electron scattering experiments~\cite{Mihovilovic:2013:ISR_exp_MAMI,Meziane:2013:PRad,Gasparian:2014:Conf:MENU,Kohl:2015:Conf:Bormio},
and muon scattering on the proton~\cite{Gilman:2013:MUSE}.
%have been proposed as well as measurement of HFS transitions.

\section{Acknowledgments}
We acknowledge support from 
the European Research Council (ERC StG.\ 279765),
the Max-Planck-Society and 
the Max-Planck-Foundation,
the Swiss National Science Foundation (project 200020-100632, 200021L\_138175, 200020\_159755) and
the Swiss Academy of Engineering Sciences, 
the BQR de l'UFR de physique fondamentale et appliqu\'ee de l'Universit\'e 
Pierre et Marie Curie- Paris 6, 
the program PAI Germaine de Sta\"el no. 07819NH du minist\`ere des affaires 
\'etrang\`eres France,
the Funda\c{c}\~{a}o para a Ci\^{e}ncia e a Tecnologia (Portugal) 
and FEDER (project 
%PTDC/FIS/82006/2006
PTDC/FIS/102110/2008
and grants SFRH/BPD/46611/2008, SFRH/BPD/74775/2010, and SFRH/BPD/76842/2011),
research centre grant No. UID/FIS/04559/2013 (LIBPhys) from FCT/MCTES/PIDDAC, Portugal,
Deutsche Forschungsgemeinschaft (DFG) GR~3172/9-1 within the D-A-CH framework,
and Ministry of Science and Technology, Taiwan,
No. 100-2112-M-007-006-MY3.
P. Indelicato acknowledges support by the 
"ExtreMe Matter Institute, Helmholtz Alliance
HA216/EMMI".

\end{document}